\documentclass[aps,prl,showpacs,preprint,superscriptaddress]{revtex4-1}

\usepackage{wrapfig}
\usepackage[]{graphicx,xcolor}
\usepackage{tabularx}
\usepackage{booktabs}
\usepackage{textcomp}
\usepackage{amsmath}
\usepackage{amssymb}
\usepackage[]{graphics}
\usepackage{amssymb}
\usepackage{amsmath}
\usepackage{ifpdf}
\ifpdf\usepackage{epstopdf}\fi

\newcommand{\RNum}[1]{\uppercase\expandafter{\romannumeral #1\relax}}
\begin{document}

\title{Ultra-Broadband Super-Planckian Radiative Heat Transfer with
  Artificial Continuum Cavity States in Patterned Hyperbolic Metamaterial}

\author{Jin~Dai}
\affiliation{Department of Applied Physics, School of Engineering
  Sciences, KTH-Royal Institute of Technology, Electrum 229, 16440
  Kista, Sweden}
\email[]{e-mail: jind@kth.se}
\author{Fei~Ding}
\affiliation{Centre for Nano Optics, University of Southern Denmark,
  Campusvej 55, DK-5230 Odense, Denmark}
\author{Sergey~I.~Bozhevolnyi}
\affiliation{Centre for Nano Optics, University of Southern Denmark,
  Campusvej 55, DK-5230 Odense, Denmark}
\author{Min~Yan}
\affiliation{Department of Applied Physics, School of Engineering
  Sciences, KTH-Royal Institute of Technology, Electrum 229, 16440
  Kista, Sweden}
\date{\today}

\begin{abstract}
Localized cavity resonances due to nanostructures at material
surfaces can greatly enhance radiative heat transfer (RHT) between two
closely placed bodies owing to stretching of cavity states in momentum
space beyond light line. Based on such understanding, we numerically
demonstrate the possibility of ultra-broadband super-Planckian
RHT between two plates patterned with
trapezoidal-shaped hyperbolic metamaterial (HMM) arrays. The
phenomenon is rooted not only in HMM's high effective index for
creating sub-wavelength resonators, but also its extremely anisotropic
isofrequency contour. The two properties enable one to create 
photonic bands  with a high spectral density to populate a desired
thermal radiation window. At sub-micron gap sizes between such two
plates, the artificial continuum states extend outside light
cone, tremendously increasing overall RHT. Our
study reveals that structured HMM offers unprecedented potential in
achieving a controllable super-Planckian radiative heat transfer
for thermal management at nanoscale.
\end{abstract}

%\keywords{Near-field, Gap Surface Plasmon, Hyperbolic, Heat transfer}

\maketitle %\maketitle must follow title, authors, abstract and \pacs

Near-field-mediated radiative heat transfer~(RHT) exceeding the
far-field blackbody limit~\cite{polder} has attracted increased
attention in recent years~\cite{RevModPhys}, not only because it
unfolds an un-explored fundamental scientific field, but also because
it holds technological importance towards nano-gap thermophotovoltaics, scanning
near-field thermal microscopy, thermal logics
etc. Commonly, materials supporting surface-guided waves at infrared
frequency were investigated for achieving such phenomenon, since surface
modes especially at frequency close to their resonances offer extra
channels of energy transfer due to evanescent-wave coupling at small
gaps. Surface-wave-bearing materials at infrared include polar
dielectrics supporting surface phonon
polaritons~\cite{RHTplate1,RHTplate2}, doped silicon supporting
surface plasmon polaritons~\cite{dopedSi}, and more recently grooved
metal surfaces supporting the so-called spoof surface plasmon
polaritons~\cite{PRBrgrating,jin1,jin2,jin3}. The presence of surface
modes leads to spectrally enhanced quasi-monochromatic heat
flux around their resonance frequencies. Contrary to the extensive
attentions paid to surface modes, \emph{localized resonant modes} were
rarely mentioned for achieving near-field-enhanced RHT. Localized
or cavity resonances at infrared or even optical frequencies can be
readily created using today's nanofabrication technologies. Spatial
localization of such modes usually corresponds to
enormous extension of the modes in momentum (wave vector $\mathbf{k}$)
dimension. Such flat bands can potentially lie outside light cone
(bounded by $k=\omega/c$) in frequency-momentum representation. Similar to
surface-mode-based RHT scenario, presence of photonic states outside light
cone implies near-field energy transfer between two
bodies, which can potentially amount to super-Planckian RHT. A direct advantage
of cavity-resonance-based RHT is that one can arrange nano-cavities
with different resonant frequencies within unit cells of two plates to
achieve enhanced RHT at multiple frequencies. With an exotic resonator
design, as the current work will reveal, an ultra-broadband
super-Planckian RHT can even be achieved.

\begin{figure}[htb!]
\centering
\includegraphics[width=0.8\columnwidth]{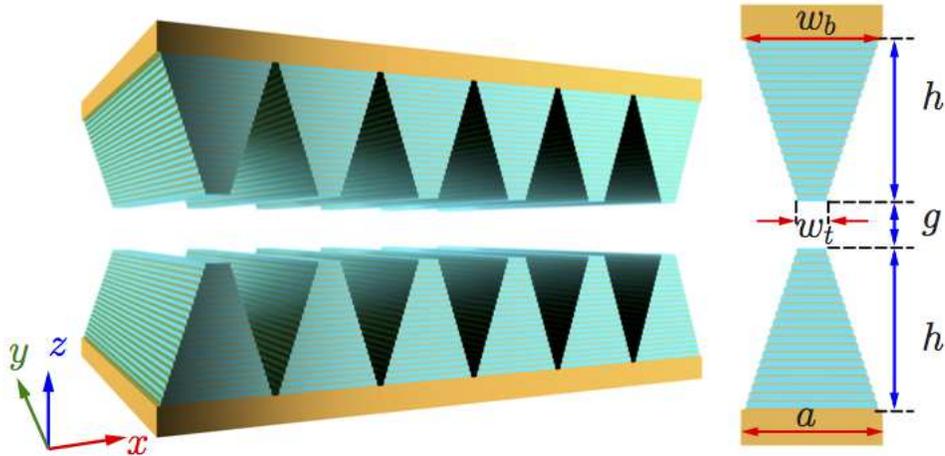}
\caption{(Color online). Schematic of the trapezoidal-shaped
  HMM plates. The cyan and orange layers denote dielectric and gold
  layers, respectively.}
\label{fig1}
\end{figure}

Referring to Fig.~\ref{fig1}, we consider RHT between two plates each
consisting a gold substrate and an array of trapezoidal-shaped
hyperbolic metamaterial (HMM). The HMM is formed by a multi-layer
metal-dielectric stack. The two plates are separated by a vacuum gap
$g$. The thicknesses of dielectric and metal are 95 and
$20$~nm, respectively. Each HMM cavity contains
20 dielectric-metal pairs. The cross-section of a single cavity
resembles a trapezoid with short base of $w_t=0.4~\mu$m, long base of
$w_b=1.9~\mu$m, and height of $h=2.3~\mu$m. The
period of the HMM arrays is fixed at $a=2.0~\mu$m. The relative
permittivities of dielectric~(Si) and metal~(Au) are
$\epsilon_{\mathrm{Si}}=11.7$ and 
$\epsilon_{\mathrm{Au}}(\omega)=1-\frac{\omega_p^2}{\omega(\omega+i\gamma)}$,
in which $\omega_p=9$~eV, and $\gamma=35$~meV, respectively. Such structure can be
fabricated with focused ion beam milling of deposited metal-dielectric
multilayers~\citep{Fei}, or with shadow deposition of dielectric and
metal layers~\cite{Yang2012,jay}.

%%%%%%%%%%% FIG 1 %%%%%%%%%%% 
%%%%%%%%%%%%%%%%%%%%%%%%%%%%

The HMM, when truncated, helps to create
\emph{subwavelength} electromagnetic cavities, while the
trapezoidal geometry is responsible for producing such resonances over
a broad wavelength range.
To simplify our argument in this paragraph, we neglect the gold
substrates\footnote{The gold substrates are mainly to prevent
  transmission leakage; their presence also induces a surface mode
  between HMM and gold. However,
the main contribution to near-field RHT is the cavity
modes, whose existence persist even without the gold substrates.} and
consider only $k_x$ wave-vector direction.
A single HMM plate, when un-patterned, has an
\emph{indefinite} diagonal effective permittivity tensor, with negative $x$ and $y$
components and positive $z$ component. Such an
anisotropic slab, for the given geometry, guides a set of $x$-propagating
$p$-polarized modes~\cite{Yan}. For frequency up to
$300\times 10^{12}~\mathrm{rad/s}$ and even higher, the modes have
\emph{almost similar} linear dispersion
curves, based on which one calculates the effective index
($n_\mathrm{eff}$) of the HMM as $\sim 3.77$. When
laterally truncated, $p$-polarized wave bounces between two 
truncation facets; each HMM patch therefore
is a 2D \emph{high-index} resonator. High-index material is
essential for creating small-dimension resonators, especially at the
upper wavelength limit for broadband RHT, that fit into
a grating period $a$. Note also that, for achieving super-Planckian
RHT at a wavelength $\lambda_r$, $a$ needs to satisfy 
$a<\lambda_r/2$ in order for the resonance to cross light line in the first Brillouin
zone of the plate's mode
spectrum. The lower wavelength limit $\lambda_\mathrm{min}$ of
desired broadband RHT would set
$a<\lambda_\mathrm{min}/2$. The most intriguing property
of such a HMM resonator, as will be further clarified in
Fig.~\ref{fig3}, is the weak dependence of its resonant frequencies
on mode orders (with nodal breaking along $z$), or equivalently
the resonator's thickness. This sets
the fundamental difference between using HMM and using an isotropic
dielectric material. A trapezoidal profile-patterned HMM can therefore
be treated as a series of vertically-stacked thin HMM resonators of varying
widths, and in turn varying resonant frequencies, for achieving
broadband operation.

We mention that a single array of such HMM resonators was
previously found to exhibit broadband absorption of far-field
radiation~\cite{cui}. Near-field properties with implications to RHT
were insofar left unexplored. There were also studies of RHT based on
un-patterned HMM
plates~\cite{MLhyper1,MLhyper2,MLhyper3,PRLHMM,Whyper,Liu:2015:metasurfaces}; the
results, as we will show later, can be radically different from RHT between
patterned HMMs, principally due to lack of localized
resonances. Here, using a rigorous full-wave scattering-matrix method~\cite{PRBrgrating,jin3},
we calculate the RHT flux between two patterned HMM plates and
numerically confirm ultra-broadband super-Planckian RHT at small gap
sizes. In addition, we utilize a finite-element based eigen-mode
solver~\cite{cwes1,cwes2} to reveal the modal
properties of the double-plate structure and identify that cavity
modes play critical roles in enhancing RHT.

The radiative heat flux between two 1D periodic arrays can be expressed by
\begin{align}
\label{eq1}
q(T_1, T_2)=\frac{1}{2\pi}\int_0^\infty[\Theta(\omega,
  T_1)-\Theta(\omega, T_2)] \Phi(\omega)\mathrm{d}\omega,
\end{align}
where  $\Theta(\omega,
T)$=$\hbar\omega/\mathrm{exp}[(\hbar\omega/k_B T) -1]$ is the mean
energy of Planck oscillators at temperature $T$ and angular frequency
$\omega$. $\Phi$ is integrated transmission factor
\begin{align}
\label{eq2}
\Phi(\omega)=\frac{1}{4\pi^2}\sum\limits_{j=s,p}\int_{-\infty}^{+\infty}\int_{-\frac{\pi}{a}}^{+\frac{\pi}{a}}\mathcal{T}_j(\omega,k_x,k_y)\mathrm{d}
k_x\mathrm{d} k_y.
\end{align}
$\mathcal{T}_j(\omega,k_x,k_y)$ is the transmission factor
that describes the probability of a thermally
excited photon transferring from one plate to the other, given
polarization $s$ or $p$, and surface-parallel wavevector
$\mathbf{k}_\parallel\equiv (k_x,k_y)$ at $\omega$.

%%%%%%%%%%%%%%%%%%%%%%%%%%%%
%%%%%%%%%%% FIG 2 %%%%%%%%%%% 
\begin{figure*}[htb!]
\centering
\includegraphics[width=0.8\columnwidth]{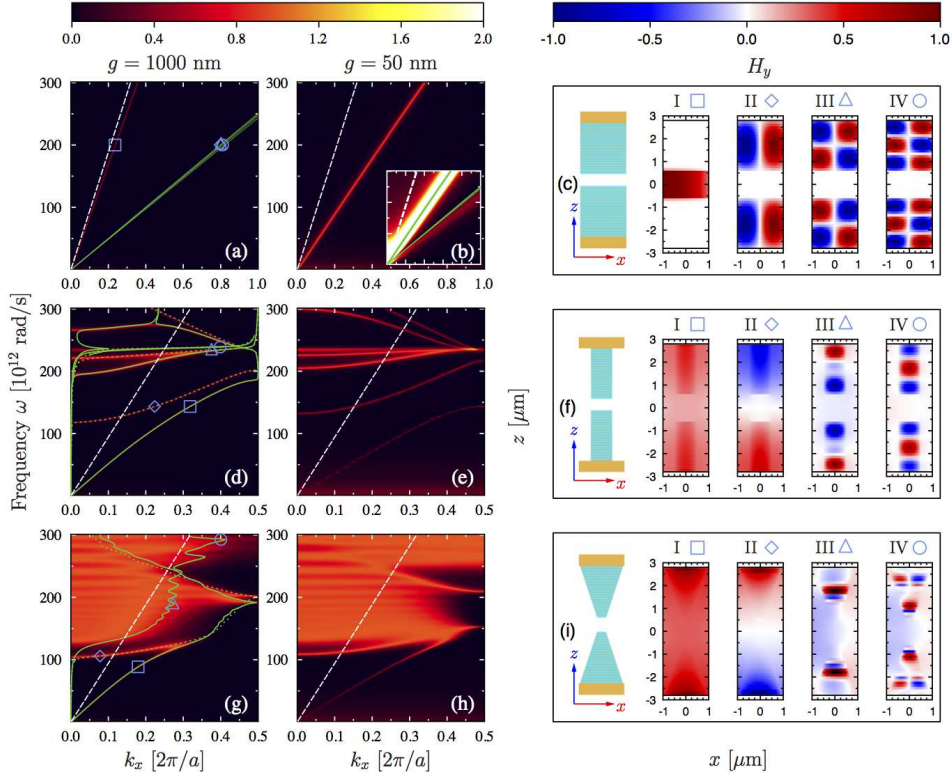}
\caption{(Color online). (a,d,g) Transmission-factor maps
  $\mathcal{T}_{s+p}(\omega,k_x,k_y=0)$ between two plates with gap $g=1000$~nm
 for three configurations: (a) unpatterned HMM plates;
(d) rectangular-profiled HMM plates $w_t=w_b=1000$~nm; (g)
  trapezoidal-profiled HMM plates. (b,e,h) The same but for $g=50$nm. The inset in (b)
has a down-limited color scale (0, 0.01). Dashed white line
  indicates light line in vacuum. Green lines are guided
  modes by the two-plate structure obtained through eigen-mode
  analysis, solid lines for bonding modes and dotted lines for
  anti-bonding modes. (c,f,g) Schematic structure units, as well as
  representative mode fields marked in (a,d,g).}
\label{fig2}
\end{figure*}

Figure~\ref{fig2} plots the transmission-factor distributions
($\mathcal{T}$ maps) over frequency and $k_x$, while $k_y$ is kept zero, 
for three types of plate configurations: unpatterned HMM plates
[panels \ref{fig2}(a) and (b)],
rectangular-profiled HMM plates  [\ref{fig2}(d)
  and (e)], and trapezoidal-profiled HMM plates
[\ref{fig2}(g) and (h)]. All configurations are
mirror-symmetric. Note the
$\mathcal{T}$ maps are shown only for $k_x$ direction, along which
the mode spectra of truncated HMM structures exhibit marked difference
against un-truncated scenario. The calculation is
repeated for two gap sizes: 1000 and 50~nm. For comparison purpose, in some
$\mathcal{T}$ maps we selectively superimpose dispersion curves
obtained from eigen-mode calculations. In Fig.~\ref{fig2}(a), when HMM
is un-truncated ($g=1000$~nm), the $\mathcal{T}$ map shows a
thin line of states just below light
line. The states originate from a gap plasmon mode
(GPM) confined mostly in the vacuum gap between the two
HMM plates, as confirmed by the mode distribution [mode I in panel
(c), or c-I] from eigen-mode analysis. Further below the GPM states,
there are a set of modes (c-II, c-III, c-IV) guided mostly in the
HMMs; the number of modes is decided by the number of metal-dielectric
layer pairs constructing the HMMs~\cite{Yan}. Only the bonding-type HMM-guided modes
are shown. Owing to their strong
confinements and the relatively large gap size, the fields in two HMM plates
are hardly coupled, therefore having almost no
contribution to the RHT process. Even when separation between the two plates is reduced to
$g=50$nm, the contribution of the HMM-guided modes to RHT is trivial,
as shown by the $\mathcal{T}$ map in Fig.~\ref{fig2}(b), as
well as its inset with a down-limited color scale. When HMM is
truncated, mode structure in the two-plate system changes
drastically. The GPM remains (f-I), but now with its field
confined between the two gold substrates; an anti-bonding GPM
(f-II) also emerge due to relatively large separation between two gold plates. 
Then, importantly, each HMM patch becomes a cavity; localized
resonances happen (modes f-III, f-IV). The cavity mode fields are
tightly confined laterally, which leads to almost flat dispersion
curves of the modes, as shown in
Fig.~\ref{fig2}(d). The contribution of these cavity modes to RHT is
evident in Fig.~\ref{fig2}(d) with $g=1000~\mathrm{nm}$, and even more
so in Fig.~\ref{fig2}(e) with $g=50~\mathrm{nm}$.

Unlike resonators made of isotropic dielectric materials, the resonant
frequencies of the HMM modes with different order numbers due to nodal breaking
along $z$ are quite close to each other. This can be understood by examining the
iso-frequency contours of the HMM in bulk, shown in $k_x$
and $k_z$ axes ($k_y=0$) in Fig.~\ref{fig3}. Three hyperbolic contours
correspond to relatively close frequencies at 235, 245, and
255$\times 10^{12}~\mathrm{rad/s}$. Given a 2D rectangular HMM cavity (of width $w$)
in Fig.~\ref{fig2}(f), resonant frequencies of modes are
decided by the modes' corresponding $k_x$ and $k_z$ values. The
fundamental mode has approximately $k_x=\pi/w$ and
$k_z=1.5\pi/h$~\footnote{Extra $0.5\pi$ phase change is due to nearly
  perfect magnetic conductor condition at boundary in contact with
  gold.}. For
next higher-order mode with a nodal breaking along $z$, $k_x$ remains
the same, and it has $k_z=2.5\pi/h$, and so on. The first three cavity
modes are indicated by three yellow dots in Fig.~\ref{fig3},
vertically aligned. They are positioned quite near to 245$~\times
10^{12}~\mathrm{rad/s}$, which is in reasonably good agreement with
Figs.~\ref{fig2}(d) and (e). The fact that the modes with nodal
breaking along $z$ stay close in frequency is fundamentally decided by
the extremely anisotropic hyperbolic iso-frequency curves of the
HMM. This is confirmed by a re-plot of Fig.~\ref{fig3} in its inset
using equal axis scales. The hyperbolic curves are almost vertically
lines.

%%%%%%%%%%% FIG 3 %%%%%%%%%%% 
\begin{figure}[htb!]
\centering
\includegraphics[width=0.8\columnwidth]{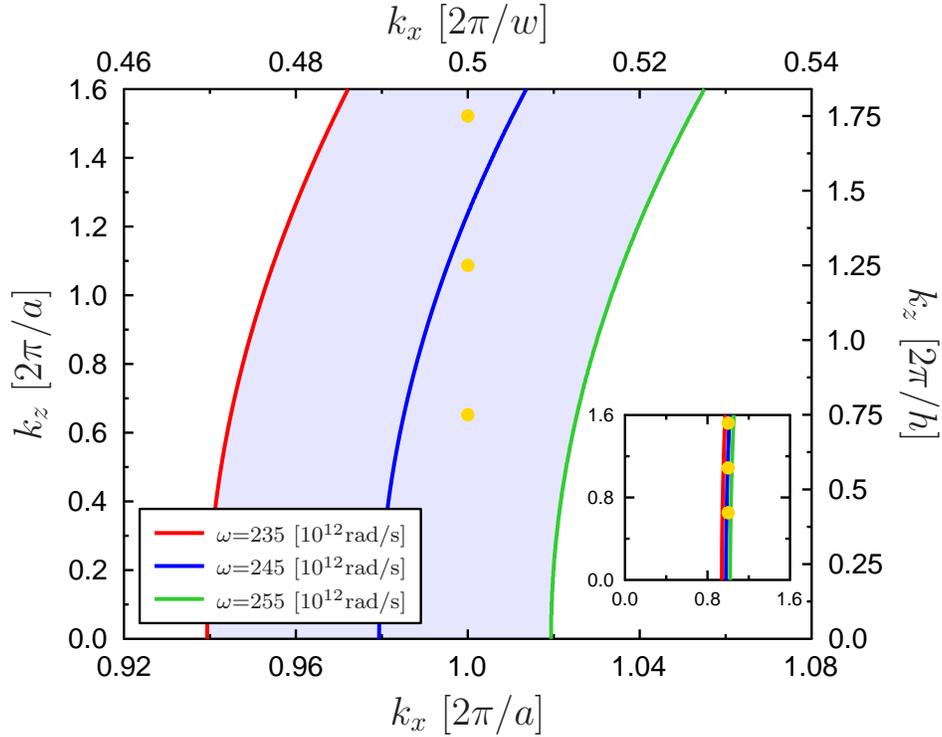}
\caption{(Color online). Iso-frequency contours of the bulk HMM
at frequencies at 235, 245, and 255$~\times
10^{12}~\mathrm{rad/s}$. The yellow dots, from bottom to top, indicate
the first three resonant modes of a rectangular HMM cavity as in
Fig.~\ref{fig2}(f). Inset shows the same plot with equal axis scales.}
\label{fig3}
\end{figure}

The iso-frequency plot in Fig.~\ref{fig3} also suggests that it is the width
of a 2D HMM cavity which determines the resonant frequencies of its
modes. The cavity height (even very thin) does not influence much the resonant
frequencies. Knowing this, one can potentially create a cavity that supports resonances at
multiple frequencies. Trapezoidal-profiled HMM structure as illustrated in
Fig.~\ref{fig1} is a straightforward solution. Indeed,
the computed $\mathcal{T}$ map for $g=1000~\mathrm{nm}$ exhibits ultra-broadband
transmission factors [Fig.~\ref{fig2}(g)]. The states contributing to
RHT are so densely packed such that they form almost a continuum. The
lower cut-off frequency of the continuum is determined by the bottom
width of the trapezoids. High transmission factors extend beyond light
line, and at certain frequencies reach the
first Brillouin zone edge. When the gap becomes smaller, at
$g=50~\mathrm{nm}$, the continuum states extend to larger $k_x$
values; more evanescent states contribute to RHT. This
lends the possibility of achieving an ultra-broadband super-Planckian RHT
between two plates using trapezoidal-profiled HMMs. From eigen-mode
calculations, we obtain modes responsible for the 
RHT process. Besides the GPM pair (modes i-I, i-II), we show modes
i-III and i-IV, which offer a glimpse of cavity
modes in the continuum. As characterized by their hot spots, the
cavity modes now have very tight $z$ confinement, or correspondingly
large $k_z$. The positions of hot spots
reveal the mechanisms of their resonances. Mode i-III
has hot spots at the middle section of the
trapezoidal HMM cavity; it well corresponds to a resonant frequency
just below $200\times 10^{12}$~rad/s. Slight complication arises at higher
frequencies. Mode i-IV, for example, has its hot spots
located at both narrow- and wide-width sections of the HMM
cavity; a wide HMM section can support a high-frequency resonance
through nodal breaking along $x$ direction.

\begin{figure}[htb!]
\centering
\includegraphics[width=0.8\columnwidth]{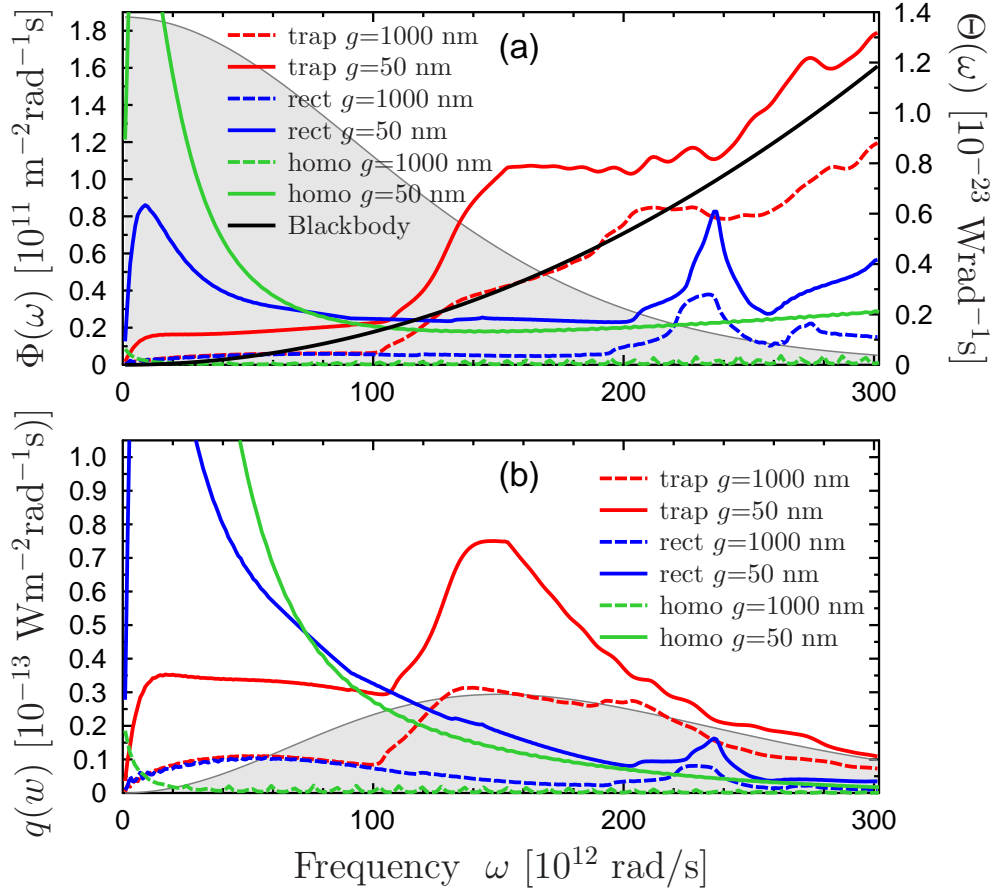}
\caption{(Color online). (a) Integrated transmission-factor spectra
  $\Phi(\omega)$ for three types of two-plate structures: trapezoidal-profiled
  (trap) HMM plates, rectangular-profiled (rect) HMM plates, and
  homogeneous (homo) HMM plates. Results for two gap sizes 1000 and
  50~nm are presented. Black line represents integrated $\Phi$ spectrum between two
  blackbodies. (b) Spectral heat flux $q(\omega)$
  for the same configurations as in (a) for plate temperatures at $301$ and
  $300$~K. The thin gray lines with shading in (a) and (b) indicate
  Planck's oscillator term $\Theta(\omega,
  \textrm{301~K})-\Theta(\omega, \textrm{300~K})$ and spectral heat
flux between two blackbodies, respectively.}
\label{fig4}
\end{figure}

A full characterization of RHT between two plates requires a
calculation of transmission factors for all
$\mathbf{k}_\parallel$ over concerned frequency range. In Supplemental Material, we
selectively plot $\mathcal{T}_{s+p}(k_x,k_y)$ maps at
$\omega=173\times 10^{12}$~rad/s for various gap
sizes. There we also explain the modal
origin of the RHT states with the help of eigen-mode analysis
especially when $k_x=0$ and $k_y\neq 0$. Generally speaking, as
$\mathbf{k}_\parallel$ deviates from $x$ direction, the near-field
contribution to RHT becomes less significant, but the far-field contribution
persists. The decreasing near-field contribution is expected
since each HMM patch no longer sustains cavity resonances when
$\mathbf{k}_\parallel$ deviates from $x$ direction. The
volumetric transmission-factor data were integrated with respect to $\mathbf{k}_x$ and $\mathbf{k}_y$. In
Fig.~\ref{fig4}(a) we plot the integrated
transmission-factor spectra, $\Phi(\omega)$, for the three types of
HMM plate configurations
as mentioned in Fig.~\ref{fig3} at gap sizes of 1000 and
50~nm. The trapezoidal-profiled HMM plates clearly exhibit the highest
$\Phi$ over almost the whole frequency range when $g=1000$~nm; at
$g=50$~nm, it has the highest $\Phi$ among the three structures at
high frequencies (above $\sim 110\times 10^{12}~\mathrm{rad/s}$). At
low frequencies, the homogeneous HMM plates have better RHT performance at
small gap sizes. That is because the effect of eddy current
generation through magnetic-field coupling ($s$ polarization) between
metal plates becomes prominent~\cite{RHTplateAu,jin3}; and the surface area
at the nearest proximity between two plates decides the degree of such
coupling (\emph{i.e.} Derjaguin proximity approximation starts to apply). The
broadband high $\Phi$ for the trapezoidal-is dominantly due to the
cavity mode continuum as shown in Figs.~\ref{fig3}(g) and (h). The
rectangular-profiled HMM plates show high $\Phi$ bands at
certain frequencies (mainly at $\sim 235\times 10^{12}$~rad/s) linked to cavity
resonances in the HMM patches, as indicated in Fig.~\ref{fig3}(d) and
(e). It is worth noticing that the peak $\Phi$ value of the rectangular-profiled HMM structure is smaller
than that of the trapezoidal-profiled HMM structure at the same
frequency. This is due to the fact that the trapezoidal one can have extra resonances due to nodal breaking in $x$
direction, while similar
high-order resonances do not exist for the rectangular counterpart in
the considered frequency range. The homogeneous HMM structure shows featureless $\Phi$
spectra, because they do not sustain any cavity resonances but
linearly dispersive guided modes [Fig.~ \ref{fig3}(a) and (b)].
As the gap size decreases down to 50~nm, the $\Phi$ spectrum of
the trapezoidal-profiled HMM structure surpasses that of
blackbody structure over the whole frequency range; ultra-broadband
super-Planckian RHT occurs. We further calculate spectral heat flux $q(\omega)$
between two plates with temperatures at 301 and 300~K for the mentioned HMM
structures [Fig.~\ref{fig4}(b)]. At $g=50$~nm, the trapezoidal
structure performs significantly better than blackbody plates, with 155\%
increase in $q$ at the highest-$q$ frequency of the blackbody-plate
scenario (\emph{i.e.}, 150$\times 10^{12}$~rad/s). Note that the
structure presented in this work is for demonstrating the flexibility of HMM-based plates
for tailoring enhanced near-field RHT [$\Phi$
spectra in Fig.~\ref{fig4}(a)] without considering specific plate
temperatures in the first place. The actual heat transfer in measurable
values [as $q$ spectra in Fig.~\ref{fig4}(b)] depends further on exact
temperature settings. In reality, towards a particular application, one should 
design the profile patterning (period, resonator width, trapezoid shape, etc) such
that one maximize near-field RHT in measurable quantities either at a desired frequency
range or over the whole spectrum.

The profile-patterning of
HMM layers is currently one dimensional, which results in
inferior RHT when $\mathbf{k}_\parallel$ has inclination towards $y$
direction. We envisage that a 2D periodic structuring of the HMM layer
(pyramid array)~\cite{Fei} can give rise to isotropic enhancement in RHT in
all $\mathbf{k}_\parallel$ directions owing to true localization of
the cavities modes. In addition, in this work we have not optimized the metal and
dielectric layer thicknesses (as well as the dielectric material type)
in the HMM. It is likely one can have even higher effective index of
the HMM by choosing appropriate geometrical and material parameters, so that one
can further adjust the frequency range of the cavity mode
continuum as well as the Brillouin zone size.

In conclusion, we numerically demonstrated one can achieve
an ultra-broadband super-Planckian RHT with two closely spaced
trapezoidal-profiled HMM plates. The design rests on two key properties
of HMMs: high effective index for creating sub-wavelength resonators,
and extremely anisotropic iso-frequency curves responsible for
cavity-width-dependent resonance frequencies.
The superiority of the trapezoidal-profiled patterning in achieving
ultra-broadband enhanced RHT is explained through its capability of
forming a cavity mode continuum. The transmission-factor maps derived from
scattering-matrix method were confirmed by dispersion curves
calculated using an eigen-mode mode solver. The contributing modes
were further elucidated with the obtained mode field
distributions. Our study reveals that highly localized cavity
resonances, besides surface waves, do enhance RHT at small separation
of two bodies. In this respect, structured hyperbolic media offers unprecedent control
in creating cavity modes for achieving a controllable super-Planckian RHT.

\begin{acknowledgments}
J. D. and M. Y. acknowledge support by the Swedish Research Council
(VR) via project 2011-4526, and VR's
Linnaeus center in Advanced Optics and Photonics. F.~D. and
S.~I.~ B. acknowledge support from the Danish Council for
Independent Research via project 1335-00104. The simulations were
performed on the Swedish National Infrastructure for Computing (SNIC).
\end{acknowledgments}

\bibliographystyle{apsrev4-1}

\begin{thebibliography}{25}%
\makeatletter
\providecommand \@ifxundefined [1]{%
 \@ifx{#1\undefined}
}%
\providecommand \@ifnum [1]{%
 \ifnum #1\expandafter \@firstoftwo
 \else \expandafter \@secondoftwo
 \fi
}%
\providecommand \@ifx [1]{%
 \ifx #1\expandafter \@firstoftwo
 \else \expandafter \@secondoftwo
 \fi
}%
\providecommand \natexlab [1]{#1}%
\providecommand \enquote  [1]{``#1''}%
\providecommand \bibnamefont  [1]{#1}%
\providecommand \bibfnamefont [1]{#1}%
\providecommand \citenamefont [1]{#1}%
\providecommand \href@noop [0]{\@secondoftwo}%
\providecommand \href [0]{\begingroup \@sanitize@url \@href}%
\providecommand \@href[1]{\@@startlink{#1}\@@href}%
\providecommand \@@href[1]{\endgroup#1\@@endlink}%
\providecommand \@sanitize@url [0]{\catcode `\\12\catcode `\$12\catcode
  `\&12\catcode `\#12\catcode `\^12\catcode `\_12\catcode `\%12\relax}%
\providecommand \@@startlink[1]{}%
\providecommand \@@endlink[0]{}%
\providecommand \url  [0]{\begingroup\@sanitize@url \@url }%
\providecommand \@url [1]{\endgroup\@href {#1}{\urlprefix }}%
\providecommand \urlprefix  [0]{URL }%
\providecommand \Eprint [0]{\href }%
\providecommand \doibase [0]{http://dx.doi.org/}%
\providecommand \selectlanguage [0]{\@gobble}%
\providecommand \bibinfo  [0]{\@secondoftwo}%
\providecommand \bibfield  [0]{\@secondoftwo}%
\providecommand \translation [1]{[#1]}%
\providecommand \BibitemOpen [0]{}%
\providecommand \bibitemStop [0]{}%
\providecommand \bibitemNoStop [0]{.\EOS\space}%
\providecommand \EOS [0]{\spacefactor3000\relax}%
\providecommand \BibitemShut  [1]{\csname bibitem#1\endcsname}%
\let\auto@bib@innerbib\@empty
%</preamble>
\bibitem [{\citenamefont {Polder}\ and\ \citenamefont {van
  Hove}(1971)}]{polder}%
  \BibitemOpen
  \bibfield  {author} {\bibinfo {author} {\bibfnamefont {D.}~\bibnamefont
  {Polder}}\ and\ \bibinfo {author} {\bibfnamefont {M.}~\bibnamefont {van
  Hove}},\ }\href {\doibase 10.1103/PhysRevB.4.3303} {\bibfield  {journal}
  {\bibinfo  {journal} {Phys. Rev. B}\ }\textbf {\bibinfo {volume} {4}},\
  \bibinfo {pages} {3303} (\bibinfo {year} {1971})}\BibitemShut {NoStop}%
\bibitem [{\citenamefont {Volokitin}\ and\ \citenamefont
  {Persson}(2007)}]{RevModPhys}%
  \BibitemOpen
  \bibfield  {author} {\bibinfo {author} {\bibfnamefont {A.~I.}\ \bibnamefont
  {Volokitin}}\ and\ \bibinfo {author} {\bibfnamefont {B.~N.~J.}\ \bibnamefont
  {Persson}},\ }\href {\doibase 10.1103/RevModPhys.79.1291} {\bibfield
  {journal} {\bibinfo  {journal} {Rev. Mod. Phys.}\ }\textbf {\bibinfo {volume}
  {79}},\ \bibinfo {pages} {1291} (\bibinfo {year} {2007})}\BibitemShut
  {NoStop}%
\bibitem [{\citenamefont {Vaillon}\ \emph {et~al.}(2010)\citenamefont
  {Vaillon}, \citenamefont {Meng{\"{u}}{\c{c}}},\ and\ \citenamefont
  {Rodolphe}}]{RHTplate1}%
  \BibitemOpen
  \bibfield  {author} {\bibinfo {author} {\bibfnamefont {M.~F.}\ \bibnamefont
  {Vaillon}}, \bibinfo {author} {\bibfnamefont {M.~P.}\ \bibnamefont
  {Meng{\"{u}}{\c{c}}}}, \ and\ \bibinfo {author} {\bibnamefont {Rodolphe}},\
  }\href {http://stacks.iop.org/0022-3727/43/i=7/a=075501} {\bibfield
  {journal} {\bibinfo  {journal} {J. Phys. D: Appl. Phys.}\ }\textbf {\bibinfo
  {volume} {43}},\ \bibinfo {pages} {75501} (\bibinfo {year}
  {2010})}\BibitemShut {NoStop}%
\bibitem [{\citenamefont {Dyakov}\ \emph {et~al.}(2014)\citenamefont {Dyakov},
  \citenamefont {Dai}, \citenamefont {Yan},\ and\ \citenamefont
  {Qiu}}]{RHTplate2}%
  \BibitemOpen
  \bibfield  {author} {\bibinfo {author} {\bibfnamefont {S.~A.}\ \bibnamefont
  {Dyakov}}, \bibinfo {author} {\bibfnamefont {J.}~\bibnamefont {Dai}},
  \bibinfo {author} {\bibfnamefont {M.}~\bibnamefont {Yan}}, \ and\ \bibinfo
  {author} {\bibfnamefont {M.}~\bibnamefont {Qiu}},\ }\href {\doibase
  10.1103/PhysRevB.90.045414} {\bibfield  {journal} {\bibinfo  {journal} {Phys.
  Rev. B}\ }\textbf {\bibinfo {volume} {90}},\ \bibinfo {pages} {045414}
  (\bibinfo {year} {2014})}\BibitemShut {NoStop}%
\bibitem [{\citenamefont {Rousseau}\ \emph {et~al.}(2009)\citenamefont
  {Rousseau}, \citenamefont {Laroche},\ and\ \citenamefont
  {Greffet}}]{dopedSi}%
  \BibitemOpen
  \bibfield  {author} {\bibinfo {author} {\bibfnamefont {E.}~\bibnamefont
  {Rousseau}}, \bibinfo {author} {\bibfnamefont {M.}~\bibnamefont {Laroche}}, \
  and\ \bibinfo {author} {\bibfnamefont {J.-J.}\ \bibnamefont {Greffet}},\
  }\href
  {http://scitation.aip.org/content/aip/journal/apl/95/23/10.1063/1.3271681}
  {\bibfield  {journal} {\bibinfo  {journal} {Appl. Phys. Lett.}\ }\textbf
  {\bibinfo {volume} {95}} (\bibinfo {year} {2009})}\BibitemShut {NoStop}%
\bibitem [{\citenamefont {Gu\'erout}\ \emph {et~al.}(2012)\citenamefont
  {Gu\'erout}, \citenamefont {Lussange}, \citenamefont {Rosa}, \citenamefont
  {Hugonin}, \citenamefont {Dalvit}, \citenamefont {Greffet}, \citenamefont
  {Lambrecht},\ and\ \citenamefont {Reynaud}}]{PRBrgrating}%
  \BibitemOpen
  \bibfield  {author} {\bibinfo {author} {\bibfnamefont {R.}~\bibnamefont
  {Gu\'erout}}, \bibinfo {author} {\bibfnamefont {J.}~\bibnamefont {Lussange}},
  \bibinfo {author} {\bibfnamefont {F.~S.~S.}\ \bibnamefont {Rosa}}, \bibinfo
  {author} {\bibfnamefont {J.-P.}\ \bibnamefont {Hugonin}}, \bibinfo {author}
  {\bibfnamefont {D.~A.~R.}\ \bibnamefont {Dalvit}}, \bibinfo {author}
  {\bibfnamefont {J.-J.}\ \bibnamefont {Greffet}}, \bibinfo {author}
  {\bibfnamefont {A.}~\bibnamefont {Lambrecht}}, \ and\ \bibinfo {author}
  {\bibfnamefont {S.}~\bibnamefont {Reynaud}},\ }\href {\doibase
  10.1103/PhysRevB.85.180301} {\bibfield  {journal} {\bibinfo  {journal} {Phys.
  Rev. B}\ }\textbf {\bibinfo {volume} {85}},\ \bibinfo {pages} {180301}
  (\bibinfo {year} {2012})}\BibitemShut {NoStop}%
\bibitem [{\citenamefont {Dai}\ \emph {et~al.}(2015)\citenamefont {Dai},
  \citenamefont {Dyakov},\ and\ \citenamefont {Yan}}]{jin1}%
  \BibitemOpen
  \bibfield  {author} {\bibinfo {author} {\bibfnamefont {J.}~\bibnamefont
  {Dai}}, \bibinfo {author} {\bibfnamefont {S.~A.}\ \bibnamefont {Dyakov}}, \
  and\ \bibinfo {author} {\bibfnamefont {M.}~\bibnamefont {Yan}},\ }\href
  {\doibase 10.1103/PhysRevB.92.035419} {\bibfield  {journal} {\bibinfo
  {journal} {Phys. Rev. B}\ }\textbf {\bibinfo {volume} {92}},\ \bibinfo
  {pages} {035419} (\bibinfo {year} {2015})}\BibitemShut {NoStop}%
\bibitem [{\citenamefont {Dai}\ \emph {et~al.}(2016{\natexlab{a}})\citenamefont
  {Dai}, \citenamefont {Dyakov},\ and\ \citenamefont {Yan}}]{jin2}%
  \BibitemOpen
  \bibfield  {author} {\bibinfo {author} {\bibfnamefont {J.}~\bibnamefont
  {Dai}}, \bibinfo {author} {\bibfnamefont {S.~A.}\ \bibnamefont {Dyakov}}, \
  and\ \bibinfo {author} {\bibfnamefont {M.}~\bibnamefont {Yan}},\ }\href
  {\doibase 10.1103/PhysRevB.93.155403} {\bibfield  {journal} {\bibinfo
  {journal} {Phys. Rev. B}\ }\textbf {\bibinfo {volume} {93}},\ \bibinfo
  {pages} {155403} (\bibinfo {year} {2016}{\natexlab{a}})}\BibitemShut
  {NoStop}%
\bibitem [{\citenamefont {Dai}\ \emph {et~al.}(2016{\natexlab{b}})\citenamefont
  {Dai}, \citenamefont {Dyakov}, \citenamefont {Bozhevolnyi},\ and\
  \citenamefont {Yan}}]{jin3}%
  \BibitemOpen
  \bibfield  {author} {\bibinfo {author} {\bibfnamefont {J.}~\bibnamefont
  {Dai}}, \bibinfo {author} {\bibfnamefont {S.~A.}\ \bibnamefont {Dyakov}},
  \bibinfo {author} {\bibfnamefont {S.~I.}\ \bibnamefont {Bozhevolnyi}}, \ and\
  \bibinfo {author} {\bibfnamefont {M.}~\bibnamefont {Yan}},\ }\href {\doibase
  10.1103/PhysRevB.94.125431} {\bibfield  {journal} {\bibinfo  {journal} {Phys.
  Rev. B}\ }\textbf {\bibinfo {volume} {94}},\ \bibinfo {pages} {125431}
  (\bibinfo {year} {2016}{\natexlab{b}})}\BibitemShut {NoStop}%
\bibitem [{\citenamefont {Ding}\ \emph {et~al.}(2014)\citenamefont {Ding},
  \citenamefont {Jin}, \citenamefont {Li}, \citenamefont {Cheng}, \citenamefont
  {Mo},\ and\ \citenamefont {He}}]{Fei}%
  \BibitemOpen
  \bibfield  {author} {\bibinfo {author} {\bibfnamefont {F.}~\bibnamefont
  {Ding}}, \bibinfo {author} {\bibfnamefont {Y.}~\bibnamefont {Jin}}, \bibinfo
  {author} {\bibfnamefont {B.}~\bibnamefont {Li}}, \bibinfo {author}
  {\bibfnamefont {H.}~\bibnamefont {Cheng}}, \bibinfo {author} {\bibfnamefont
  {L.}~\bibnamefont {Mo}}, \ and\ \bibinfo {author} {\bibfnamefont
  {S.}~\bibnamefont {He}},\ }\href {\doibase 10.1002/lpor.201400157} {\bibfield
   {journal} {\bibinfo  {journal} {Laser Photon. Rev.}\ }\textbf {\bibinfo
  {volume} {8}},\ \bibinfo {pages} {946} (\bibinfo {year} {2014})}\BibitemShut
  {NoStop}%
\bibitem [{\citenamefont {Yang}\ \emph {et~al.}(2012)\citenamefont {Yang},
  \citenamefont {Yao}, \citenamefont {Rho}, \citenamefont {Yin},\ and\
  \citenamefont {Zhang}}]{Yang2012}%
  \BibitemOpen
  \bibfield  {author} {\bibinfo {author} {\bibfnamefont {X.}~\bibnamefont
  {Yang}}, \bibinfo {author} {\bibfnamefont {J.}~\bibnamefont {Yao}}, \bibinfo
  {author} {\bibfnamefont {J.}~\bibnamefont {Rho}}, \bibinfo {author}
  {\bibfnamefont {X.}~\bibnamefont {Yin}}, \ and\ \bibinfo {author}
  {\bibfnamefont {X.}~\bibnamefont {Zhang}},\ }\href {\doibase
  10.1038/nphoton.2012.124} {\bibfield  {journal} {\bibinfo  {journal} {Nat.
  Photon.}\ }\textbf {\bibinfo {volume} {6}},\ \bibinfo {pages} {450} (\bibinfo
  {year} {2012})}\BibitemShut {NoStop}%
\bibitem [{\citenamefont {Zhou}\ \emph {et~al.}(2014)\citenamefont {Zhou},
  \citenamefont {Kaplan}, \citenamefont {Chen},\ and\ \citenamefont
  {Guo}}]{jay}%
  \BibitemOpen
  \bibfield  {author} {\bibinfo {author} {\bibfnamefont {J.}~\bibnamefont
  {Zhou}}, \bibinfo {author} {\bibfnamefont {A.~F.}\ \bibnamefont {Kaplan}},
  \bibinfo {author} {\bibfnamefont {L.}~\bibnamefont {Chen}}, \ and\ \bibinfo
  {author} {\bibfnamefont {L.~J.}\ \bibnamefont {Guo}},\ }\href {\doibase
  10.1021/ph5001007} {\bibfield  {journal} {\bibinfo  {journal} {ACS
  Photonics}\ }\textbf {\bibinfo {volume} {1}},\ \bibinfo {pages} {618}
  (\bibinfo {year} {2014})}\BibitemShut {NoStop}%
\bibitem [{Note1()}]{Note1}%
  \BibitemOpen
  \bibinfo {note} {The gold substrates are mainly to prevent transmission
  leakage; their presence also induces a surface mode between HMM and gold.
  However, the main contribution to near-field RHT is the cavity modes, whose
  existence persist even without the gold substrates.}\BibitemShut {Stop}%
\bibitem [{\citenamefont {Yan}\ \emph {et~al.}(2011)\citenamefont {Yan},
  \citenamefont {Thyl\'{e}n},\ and\ \citenamefont {Qiu}}]{Yan}%
  \BibitemOpen
  \bibfield  {author} {\bibinfo {author} {\bibfnamefont {M.}~\bibnamefont
  {Yan}}, \bibinfo {author} {\bibfnamefont {L.}~\bibnamefont {Thyl\'{e}n}}, \
  and\ \bibinfo {author} {\bibfnamefont {M.}~\bibnamefont {Qiu}},\ }\href
  {\doibase 10.1364/OE.19.003818} {\bibfield  {journal} {\bibinfo  {journal}
  {Opt. Express}\ }\textbf {\bibinfo {volume} {19}},\ \bibinfo {pages} {3818}
  (\bibinfo {year} {2011})}\BibitemShut {NoStop}%
\bibitem [{\citenamefont {Cui}\ \emph {et~al.}(2012)\citenamefont {Cui},
  \citenamefont {Fung}, \citenamefont {Xu}, \citenamefont {Ma}, \citenamefont
  {Jin}, \citenamefont {He},\ and\ \citenamefont {Fang}}]{cui}%
  \BibitemOpen
  \bibfield  {author} {\bibinfo {author} {\bibfnamefont {Y.}~\bibnamefont
  {Cui}}, \bibinfo {author} {\bibfnamefont {K.~H.}\ \bibnamefont {Fung}},
  \bibinfo {author} {\bibfnamefont {J.}~\bibnamefont {Xu}}, \bibinfo {author}
  {\bibfnamefont {H.}~\bibnamefont {Ma}}, \bibinfo {author} {\bibfnamefont
  {Y.}~\bibnamefont {Jin}}, \bibinfo {author} {\bibfnamefont {S.}~\bibnamefont
  {He}}, \ and\ \bibinfo {author} {\bibfnamefont {N.~X.}\ \bibnamefont
  {Fang}},\ }\href {\doibase 10.1021/nl204118h} {\bibfield  {journal} {\bibinfo
   {journal} {Nano Letters}\ }\textbf {\bibinfo {volume} {12}},\ \bibinfo
  {pages} {1443} (\bibinfo {year} {2012})}\BibitemShut {NoStop}%
\bibitem [{\citenamefont {Biehs}\ \emph {et~al.}(2013)\citenamefont {Biehs},
  \citenamefont {Tschikin}, \citenamefont {Messina},\ and\ \citenamefont
  {Ben-Abdallah}}]{MLhyper1}%
  \BibitemOpen
  \bibfield  {author} {\bibinfo {author} {\bibfnamefont {S.-A.}\ \bibnamefont
  {Biehs}}, \bibinfo {author} {\bibfnamefont {M.}~\bibnamefont {Tschikin}},
  \bibinfo {author} {\bibfnamefont {R.}~\bibnamefont {Messina}}, \ and\
  \bibinfo {author} {\bibfnamefont {P.}~\bibnamefont {Ben-Abdallah}},\ }\href
  {\doibase http://dx.doi.org/10.1063/1.4800233} {\bibfield  {journal}
  {\bibinfo  {journal} {Appl. Phys. Lett.}\ }\textbf {\bibinfo {volume}
  {102}},\ \bibinfo {pages} {131106} (\bibinfo {year} {2013})}\BibitemShut
  {NoStop}%
\bibitem [{\citenamefont {Guo}\ and\ \citenamefont {Jacob}(2013)}]{MLhyper2}%
  \BibitemOpen
  \bibfield  {author} {\bibinfo {author} {\bibfnamefont {Y.}~\bibnamefont
  {Guo}}\ and\ \bibinfo {author} {\bibfnamefont {Z.}~\bibnamefont {Jacob}},\
  }\href {\doibase 10.1364/OE.21.015014} {\bibfield  {journal} {\bibinfo
  {journal} {Opt. Express}\ }\textbf {\bibinfo {volume} {21}},\ \bibinfo
  {pages} {15014} (\bibinfo {year} {2013})}\BibitemShut {NoStop}%
\bibitem [{\citenamefont {Guo}\ \emph {et~al.}(2012)\citenamefont {Guo},
  \citenamefont {Cortes}, \citenamefont {Molesky},\ and\ \citenamefont
  {Jacob}}]{MLhyper3}%
  \BibitemOpen
  \bibfield  {author} {\bibinfo {author} {\bibfnamefont {Y.}~\bibnamefont
  {Guo}}, \bibinfo {author} {\bibfnamefont {C.~L.}\ \bibnamefont {Cortes}},
  \bibinfo {author} {\bibfnamefont {S.}~\bibnamefont {Molesky}}, \ and\
  \bibinfo {author} {\bibfnamefont {Z.}~\bibnamefont {Jacob}},\ }\href
  {\doibase http://dx.doi.org/10.1063/1.4754616} {\bibfield  {journal}
  {\bibinfo  {journal} {Appl. Phys. Lett.}\ }\textbf {\bibinfo {volume}
  {101}},\ \bibinfo {pages} {131106} (\bibinfo {year} {2012})}\BibitemShut
  {NoStop}%
\bibitem [{\citenamefont {Miller}\ \emph {et~al.}(2014)\citenamefont {Miller},
  \citenamefont {Johnson},\ and\ \citenamefont {Rodriguez}}]{PRLHMM}%
  \BibitemOpen
  \bibfield  {author} {\bibinfo {author} {\bibfnamefont {O.~D.}\ \bibnamefont
  {Miller}}, \bibinfo {author} {\bibfnamefont {S.~G.}\ \bibnamefont {Johnson}},
  \ and\ \bibinfo {author} {\bibfnamefont {A.~W.}\ \bibnamefont {Rodriguez}},\
  }\href {\doibase 10.1103/PhysRevLett.112.157402} {\bibfield  {journal}
  {\bibinfo  {journal} {Phys. Rev. Lett.}\ }\textbf {\bibinfo {volume} {112}},\
  \bibinfo {pages} {157402} (\bibinfo {year} {2014})}\BibitemShut {NoStop}%
\bibitem [{\citenamefont {Biehs}\ \emph {et~al.}(2012)\citenamefont {Biehs},
  \citenamefont {Tschikin},\ and\ \citenamefont {Ben-Abdallah}}]{Whyper}%
  \BibitemOpen
  \bibfield  {author} {\bibinfo {author} {\bibfnamefont {S.-A.}\ \bibnamefont
  {Biehs}}, \bibinfo {author} {\bibfnamefont {M.}~\bibnamefont {Tschikin}}, \
  and\ \bibinfo {author} {\bibfnamefont {P.}~\bibnamefont {Ben-Abdallah}},\
  }\href {\doibase 10.1103/PhysRevLett.109.104301} {\bibfield  {journal}
  {\bibinfo  {journal} {Phys. Rev. Lett.}\ }\textbf {\bibinfo {volume} {109}},\
  \bibinfo {pages} {104301} (\bibinfo {year} {2012})}\BibitemShut {NoStop}%
\bibitem [{\citenamefont {Liu}\ and\ \citenamefont
  {Zhang}(2015)}]{Liu:2015:metasurfaces}%
  \BibitemOpen
  \bibfield  {author} {\bibinfo {author} {\bibfnamefont {X.}~\bibnamefont
  {Liu}}\ and\ \bibinfo {author} {\bibfnamefont {Z.}~\bibnamefont {Zhang}},\
  }\href@noop {} {\bibfield  {journal} {\bibinfo  {journal} {ACS Photonics}\
  }\textbf {\bibinfo {volume} {2}},\ \bibinfo {pages} {1320} (\bibinfo {year}
  {2015})}\BibitemShut {NoStop}%
\bibitem [{\citenamefont {Parisi}\ \emph {et~al.}(2012)\citenamefont {Parisi},
  \citenamefont {Zilio},\ and\ \citenamefont {Romanato}}]{cwes1}%
  \BibitemOpen
  \bibfield  {author} {\bibinfo {author} {\bibfnamefont {G.}~\bibnamefont
  {Parisi}}, \bibinfo {author} {\bibfnamefont {P.}~\bibnamefont {Zilio}}, \
  and\ \bibinfo {author} {\bibfnamefont {F.}~\bibnamefont {Romanato}},\ }\href
  {\doibase 10.1364/OE.20.016690} {\bibfield  {journal} {\bibinfo  {journal}
  {Opt. Express}\ }\textbf {\bibinfo {volume} {20}},\ \bibinfo {pages} {16690}
  (\bibinfo {year} {2012})}\BibitemShut {NoStop}%
\bibitem [{\citenamefont {Fietz}\ \emph {et~al.}(2011)\citenamefont {Fietz},
  \citenamefont {Urzhumov},\ and\ \citenamefont {Shvets}}]{cwes2}%
  \BibitemOpen
  \bibfield  {author} {\bibinfo {author} {\bibfnamefont {C.}~\bibnamefont
  {Fietz}}, \bibinfo {author} {\bibfnamefont {Y.}~\bibnamefont {Urzhumov}}, \
  and\ \bibinfo {author} {\bibfnamefont {G.}~\bibnamefont {Shvets}},\ }\href
  {\doibase 10.1364/OE.19.019027} {\bibfield  {journal} {\bibinfo  {journal}
  {Opt. Express}\ }\textbf {\bibinfo {volume} {19}},\ \bibinfo {pages} {19027}
  (\bibinfo {year} {2011})}\BibitemShut {NoStop}%
\bibitem [{Note2()}]{Note2}%
  \BibitemOpen
  \bibinfo {note} {Extra $0.5\pi $ phase change is due to nearly perfect
  magnetic conductor condition at boundary in contact with gold.}\BibitemShut
  {Stop}%
\bibitem [{\citenamefont {Chapuis}\ \emph {et~al.}(2008)\citenamefont
  {Chapuis}, \citenamefont {Volz}, \citenamefont {Henkel}, \citenamefont
  {Joulain},\ and\ \citenamefont {Greffet}}]{RHTplateAu}%
  \BibitemOpen
  \bibfield  {author} {\bibinfo {author} {\bibfnamefont {P.-O.}\ \bibnamefont
  {Chapuis}}, \bibinfo {author} {\bibfnamefont {S.}~\bibnamefont {Volz}},
  \bibinfo {author} {\bibfnamefont {C.}~\bibnamefont {Henkel}}, \bibinfo
  {author} {\bibfnamefont {K.}~\bibnamefont {Joulain}}, \ and\ \bibinfo
  {author} {\bibfnamefont {J.-J.}\ \bibnamefont {Greffet}},\ }\href {\doibase
  10.1103/PhysRevB.77.035431} {\bibfield  {journal} {\bibinfo  {journal} {Phys.
  Rev. B}\ }\textbf {\bibinfo {volume} {77}},\ \bibinfo {pages} {035431}
  (\bibinfo {year} {2008})}\BibitemShut {NoStop}%
\end{thebibliography}

\newpage
\begin{center}
\textbf{Supplemental Material}
\end{center}
\begin{figure}[htb!]
\centering
\includegraphics[width=0.6\columnwidth]{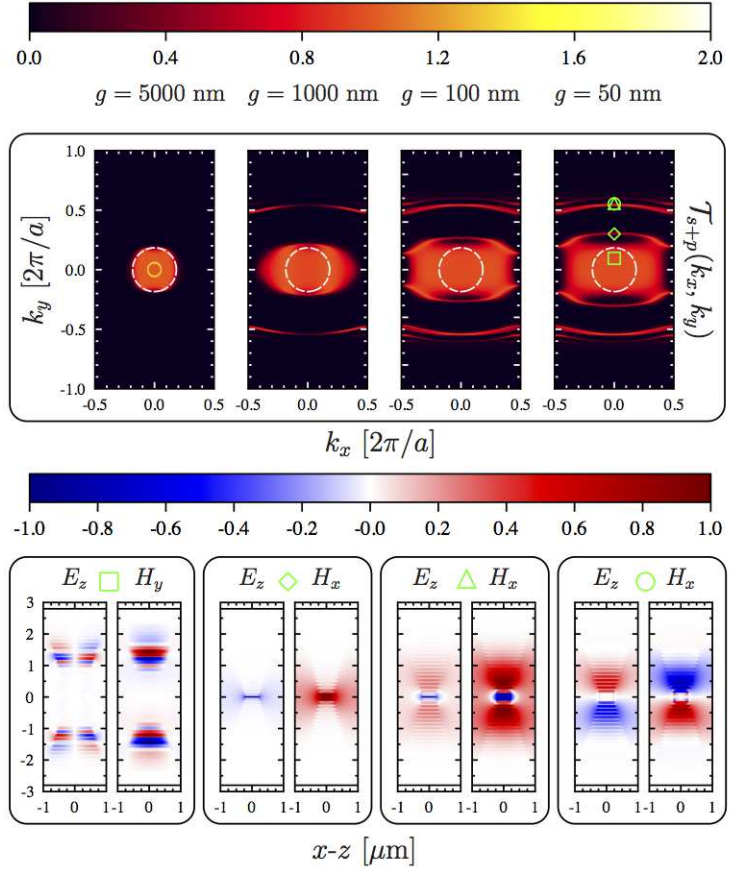}
\caption{(Color online). Upper panels show transmission factor maps
  $\mathcal{T}_{s+p}(k_x,k_y)$ between the two
  trapezoidal-profiled HMM plates at four gap sizes at
  $\omega=173\times 10^{12}$~rad/s. Both polarizations are
    included. Circles in dashed white lines are
    light cones. Lower panels present four modes in their
    respective major electric- and magnetic-field components at various $k_y$
    values ($k_x=0$) for the structure with $g=50$~nm.}
\label{fig_s1}
\end{figure}

Figure~\ref{fig_s1} shows transmission-factor ($\mathcal{T}$) maps
as a function of surface-parallel wavevectors ($k_x,k_y$) at a fixed frequency
$\omega=173\times 10^{12}$~rad/s for the trapezoidal-profiled HMM plate
structure discussed in the main text. Four scenarios corresponding to
gap sizes of 5000, 1000, 100, and 50~nm are shown. Integration of
each spectra over $k_x$ and $k_y$ gives rise to the integrated
transmission-factor $\Phi$ at this frequency.

From the upper panels in Fig.~\ref{fig_s1}, one sees that at larger gap sizes the
contribution to radiative heat transfer (RHT) mainly comes from
electromagnetic states inside light cone. These states correspond to
thermally excited photons
radiating away from one plate to the other. It is interesting to
notice that the combined $s+p$ states are almost isotropic in the
$(k_x,k_y)$ plane; they form nearly unitary transmission factors
filling the whole light cone. The amplitude of the transmission
factors are still less than those between two ideal blackbody plates,
which would have a uniform amplitude of two filling the whole light cone.
As the gap size decreases, states outside the light
cone come into play, suggesting more and more near-field
contributions to RHT. At $g=1000$~nm, the integrated transmission
factor at this frequency is 
already beyond that between the far-field blackbody limit~[see Fig.~4(a) in the main article].

Besides the main block of RHT states connected to light cone, there
are discrete thin lines of states emerging and their contributions
become more significant as gap size decreases. To better understand
these RHT channels, in the lower panels in Fig.~\ref{fig_s1} we plot four
representative mode fields as marked on the $\mathcal{T}$ spectrum for
the $g=50$~nm configuration. The modes were 
calculated using a finite-element based eigen-mode solver. We examine
particularly modes with $(k_x=0,k_y\neq~0$). Mode patterns for
($k_x\neq~0,k_y=0$) were presented in Fig.~2 in the main text. The
modes in Fig.~\ref{fig_s1} are nothing but guided modes by the
trapezoidal-profiled HMM plates along $y$ direction. Each trapezoidal
HMM stack, being structurally invariant in $y$ direction, functions
like an electromagnetic waveguide; a single grating with periodic
arrangement of such HMM stacks is a waveguide
array. Due to high contrast in permittivity values between HMM and
vacuum, the guided modes are hybrid in polarization. the
mode inside the light cone (marked by green square) is a radiation mode, which
is manifested by its rapid variation in mode profile plotted in
Fig.~\ref{fig_s1}. The modes outside light cone are in principle
similar to those presented for the un-patterned HMM plate structure in Fig.~2(c) in the main article. The mode marked by the diamond is a gap
plasmon mode mainly confined inside the air gap between the two HMM
stacks. The modes marked by the 
triangle and the circle are a bonding- and anti-bonding mode pair
whose fields are mainly confined in HMMs. Note that, these two
HMM-guided modes are the first two (fundamental mode pair), among a set of such
HMM-guided modes supported by the system.

\end{document}